\documentclass[letter]{aa}  
\usepackage{graphicx}
\usepackage{txfonts}
\usepackage{epsfig}

\def\mso{\,{\rm M}_\odot}

 \def\kms{\, {\rm km}\, {\rm s}^{-1}}

 \def\simle{\mathrel{\hbox{\rlap{\hbox{\lower4pt\hbox{$\sim$}}}\hbox{$<$}}}}
 \def\simgr{\mathrel{\hbox{\rlap{\hbox{\lower4pt\hbox{$\sim$}}}\hbox{$>$}}}}
 \def\msoy{\, \mso~{\rm yr}^{-1}}

 \def\erg{\, {\rm erg}}
 \def\gcms{\,{\rm g\,cm}\,{\rm s}^{-1}}

\begin{document}
   \title{Multiple ring nebulae around blue supergiants}


   \author{S. M. Chi\c{t}\v{a}
          \inst{1}
          \and
          N. Langer
	  \inst{1}
          \and 
	  A. J. van Marle
	  \inst{1,2}
          \and
          G. Garc\'{i}a-Segura
          \inst{3} 
           \and
           A. Heger
           \inst{4,5}}

   \offprints{S. M. Chi\c{t}\v{a}, s.m.chita@uu.nl}

   \institute{\inst{1}
              Astronomical Institute, Utrecht University,
              PO Box 80000, 3508 TA Utrecht, The Netherlands\\
              \inst{2}
               Bartol Research Institute,University of Delaware,
               102 Sharp laboratory, Newark, 19716 DE, USA \\
              \inst{3}
               Instituto de Astronom\'{i}a - UNAM,
               APDO Postal 877, Ensenada, 22800 Baja California, Mexico\\
              \inst{4}
               Theoretical Astrophysics Group,
               T-6,MS B227 Los Alamos National Laboratory, Los Alamos NM 87545, USA\\
               \inst{5}
	       School of Physics and Astronomy 112 Church st, 
	       University of Minnesota, Minneapolis, MN 55455, USA\\
	       }

   \date{Received 29 april 2008; accepted 16 july 2008}

 
  \abstract
  {In the course of the life of a massive star, wind-wind interaction can give
   rise to the formation of circumstellar nebulae which are both predicted and observed in the nature.}
  {We present generic model calculations to predict the properties of such nebulae for blue supergiants.}
  {From stellar evolution calculations including rotation, we obtain the time dependence
  of the stellar wind properties and of the stellar radiation field. These are used
  as input for hydro-calculations of the circumstellar medium throughout the star's life.}
  {Here, we present the results for a rapidly rotating 12$\mso$ single star. This star undergoes a blue loop
   during its post main sequence evolution, at the onset of which its contraction spins it up 
   close to critical rotation. Due to the consequent anisotropic mass loss, 
   the blue supergiant wind sweeps up
   the preceding slow wind into an hour glass structure. Its collision with
   the previously formed spherical red supergiant wind shell forms a short-lived
   luminous nebula consisting of two polar caps and a central inner ring. 
   With time, the polar caps evolve into mid-latitude
   rings which gradually move toward the equatorial plane while the central ring is fading.
   These structures are reminiscent to the observed nebulae around the blue supergiant 
   Sher~25 and the progenitor of SN~1987A. }
   {The simple model of an hour glass colliding with a spherical shell retrieves most of the
    intriguing nebula geometries discovered around blue supergiants, and suggests them to form
    an evolutionary sequence. Our results indicate that binarity is not required to obtain them.}

   \keywords{-hydrodynamics - ISM: bubbles - Stars: winds,outflows - Stars: supergiants}

   \authorrunning{S. M. Chi\c{t}\v{a} et al.}
   \titlerunning{Multiple ring nebulae around blue supergiants}
   \maketitle
%

\section{Introduction}
 
During the course of their evolution, massive stars have strong winds
which eject matter into their surroundings. During their post-main sequence
evolution, these stars can move back and forth from the blue to the red side
of the Hertzsprung-Russell (HR)~diagram and back to the red, with little time spent at intermediate
effective temperatures (e.g., Langer~{\cite{L91b}}). Hydrodynamic considerations
imply that each such transition does produce a circumstellar shell: 
When the star moves from the blue to the
red side of the HR~diagram, the slow red supergiant (RSG) wind will be stalled by the
high pressure of the previously created hot wind bubble, and will accumulate into a
shell at the location where this pressure equals the RSG wind ram pressure
(Garc\'{i}a-Segura et al.~{\cite{GLM96b}}). We call such a more or less stationary shell the RSG~shell. 
When the star moves from the red to the blue
side of the HR~diagram, the wind speed increases and the blue supergiant (BSG) wind
plows up the preceding RSG wind into a rapidly expanding shell, which we call the BSG~shell. 
 
Consequently, we expect a spectacular circumstellar phenomenon
for stars undergoing so called blue loops, namely that it triggers
the formation of an expanding BSG~shell, which 
will at some point smash into the previously formed stationary RSG~shell.  
While both, the RSG and the BSG~shell by itself, may be difficult to observe,
their violent interaction may release enough energy to provide an observable
nebula. 

Despite this simple and intriguing expectation, there are so far only
few attempts to obtain quantitative prediction for the outcome of the 
described shell interaction (see Blondin et al.~{\cite{BL93}}, 
Martin et al.~{\cite{MA95}}, Podsiadlowski et al.~{\cite{PM05}}).
Within an effort to describe 
this phenomenon through generic calculations, which use detailed stellar
evolution models as input for the circumstellar hydrodynamic modeling 
(Chi\c{t}\v{a} et. al., in preparation), we focus here on the results for a 
rotating 12$\mso$ single star.

\section{Computational method}

As input for our circumstellar hydrodynamic calculations, we use the
results of a stellar evolution calculation for a star of 12~$\mso$ 
and a metallicity of $Z=0.02$. Specifically, we utilize Model F12B from 
Heger \& Langer~({\cite{HL00}}), which has an initial rotational velocity of 328~$\kms$.
The code used to compute this model includes OPAL opacities,
detailed nuclear networks, mass loss according to Nieuwenhuijzen \& Jager~({\cite{NJ90}}), 
the physics of rotation
for the stellar interior, and rotationally modulated stellar winds,
as described in Heger, Langer \& Woosley~({\cite{HLW00}}).

\begin{table}
\caption{Ejected mass ($\Delta M$), momentum ($\Delta p$) and kinetic energy
($\Delta E$) during the various evolutionary phases of our stellar model. The evolutionary phase 
is identified
in the first column: main sequence phase~(MS), first red supergiant phase~(RSGI), 
phase of rapid rotation~(RR), blue supergiant stage~(BSG), and second red supergiant phase~(RSGII).
The approximate duration of each phase is given in the second column. }
\label{table:F12B}      
\centering                          
\begin{tabular}{c c c c c}        
\hline
Phase  & $\Delta$t   & $\Delta$M & $\Delta$p      & $\Delta$E      \\
       & $10^3\,$yr  & $\mso$    & $10^{38}\gcms$ & $10^{45}\,$erg \\
\hline
MS     & 19200       & 0.43      & 396        & 1480 \\
RSG~I  & 825         & 0.33      & 71         & 38   \\
RR     & 25          & 0.02      & 7.2          & 6.0    \\
BSG    & 550         & 0.11      & 52         & 68   \\
RSG~II & 225         & 0.13      & 25         & 12   \\
\hline
\end{tabular}
\end{table}

The evolution of the stellar model in the HR diagram is show in
Fig.~\ref{FigHrd}. At core-H exhaustion, it moves to the RSG regime
where it remains for 825\,000 yrs ($\sim$60~\% of the core-He burning life time),
before it undergoes a blue loop. It then stays in the BSG regime of the
HR~diagram for the remaining part of core helium burning, before it
moves back to the RSG regime where it explodes as a Type~II supernova.
 
As shown by Heger \& Langer~({\cite{HL98}}), as the convective envelope 
retreats during the onset of the blue loop, all its angular momentum
is concentrated in a small amount of mass in the top layers of the star
by the time convection vanishes. Blue loops therefore provide a natural
way to bring the stellar surface to close to critical rotation. This does also
happen in our chosen stellar model (Fig.~\ref{FigMvvo}). The limit of critical
rotation is reached during the red-blue transition, which produces 
a brief period of strong, slow and anisotropic mass loss (Table~1).
The strong mass loss then reduces the rotation rate of the stellar surface
(Langer~{\cite{L98}}), and the star settles at a rotation velocity of
about~50$\kms$ in the BSG regime.

To simulate the evolution of the circumstellar matter (CSM) 
we use the ZEUS~3D code developed by 
Stone \& Norman~({\cite{SN92}}) and Clark~({\cite{Clk96}}).
ZEUS~3D is an explicit non-conservative code that solves the hydrodynamical equations as partial, finite 
difference equations on a fixed, staggered mesh grid.
Radiatively optically-thin cooling is included by
solving the energy equation implicitly according to Mac Low et al.~({\cite{M89}}),
and by using the plasma cooling curve of MacDonald \& Bailey~({\cite{M81}}).
We compute the evolution of the CSM during the main sequence and the early RSG stage in 1D, 
with 4500 grid points over a radius of 45~pc, 
and we assume an interstellar medium density of 1~cm$^{-3}$. 
After 100\,000\,yr into the first RSG stage, 
we map the 1D model onto a 2D spherical grid to compute its further evolution. 
The inflow inner boundary condition is applied at 0.025~pc, and the outer boundary remained at 45~pc. 
The radial component of the grid is resolved with 1000 grid points, where 900 grid points are used
for the inner 5~pc, and 100 grid points for the outer 40~pc. 
The angular coordinate of 90 degrees is resolved with 200 grid points.
The method used here was applied before by Garc\'{i}a-Segura et al.~({\cite{GF96}}, 
~{\cite{GML96a}} and ~{\cite{GLM96b}}) and 
van Marle et al.~({\cite{MLG05}} and ~{\cite{MLG07}}).  

We are using the time dependent mass loss rate and the terminal wind speed from the
stellar evolution model as input in our central mesh point for the
hydrodynamic calculations. The wind speed is obtained from the
stellar escape velocity using the scaling proposed by Eldridge~({\cite{E05}}).
The wind anisotropy is described using the equations of Bjorkman \&
Cassinelli~({\cite{BC93}}), as in Langer et al.~({\cite{LGM99}}).
For near-critically rotating stars, this provides a slow and dense equatorial
outflow, and a fast wind in polar directions.
We note that while the Bjorkman-Cassinelli mechanism has been criticized
in the context of line driven winds (Owocki et al.~{\cite{O96}}), it is
unclear whether line driving does play a major role in the situation of 
near-critical rotation.

The effect of photoionization was included in the simulations
by calculating the Str\"{o}mgren radius along each radial
grid line as described in Garc\'{i}a-Segura et al.~
({\cite{GLRF99}}) and van Marle et al.~({\cite{MLG05}},
~{\cite{MLG07}} and ~{\cite{MLYG08}}).
The number of ionizing photons is computed according to the
effective temperatures and surface gravities of the stellar evolution
model, by interpolating in a grid of model
atmospheres for massive OB~stars of solar metallicity computed with the
FASTWIND non-LTE code (Puls et al.~{\cite{Pu05}})
as described in Lefever et al.~({\cite{Le07}}).

\section{Results}

\begin{figure*}[t]
   \centering
   \includegraphics[width=0.65 \textwidth, angle=270]{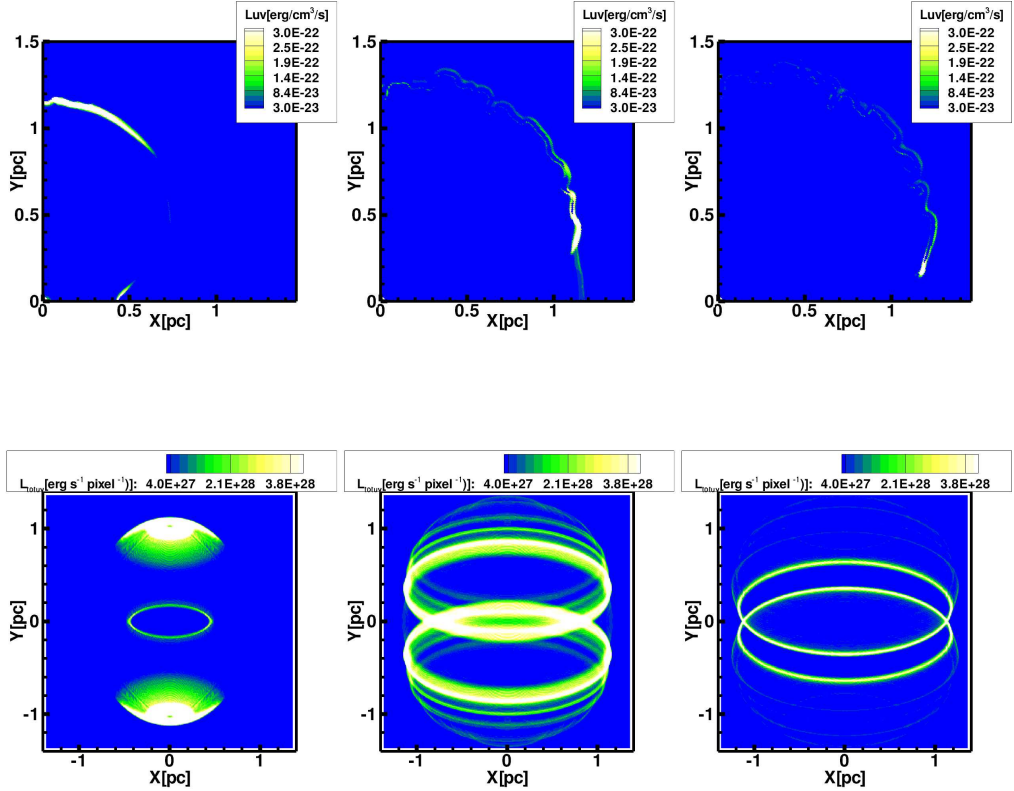}
      \caption{Emission structures from our 2D hydro simulation, for the same
moments in time as panels 2 to 4 of Fig\ref{FigDenstemp}, i.e.,
9000\,yr, 15\,000\,yr, and 18\,000\,yr after the onset of the BSG wind.  
{\bf Upper panel}: Emissivity of the gas with $10^{4} < T < 10^{6}\,$K,
in the simulation plane. 
{\bf Lower panel}: Projections of the 3D emission obtained by assuming
rotational symmetry of the 2D structures of the upper panel, viewed with
an inclination angle of 60$^{\circ}$, constructed with a resolution of
400$\times$400 pixels. }
       \label{FigEmis}
   \end{figure*} 

During its main sequence phase our 12$\mso$ star creates a
hot bubble in the interstellar medium which, at core
hydrogen exhaustion, is characterized by a radius of
30\,pc and an internal energy density of $10^{-12}\erg$\,cm$^{-3}$.
 Once the star has become a RSG, a slow ($\sim 50\kms$),
dense and isotropic wind is injected into the computational domain (Fig.~\ref{FigMvvo}).
This RSG wind accumulates at a distance of $\sim 1.5\,$pc
where its ram pressure is balanced by the hot bubble pressure, and forms
a RSG shell (cf. Garc\'{i}a-Segura et al.~{\cite{GLM96b}}). At the end of the first RSG phase,
this shell contains about $0.26\mso$.  
It is rather extended ($\Delta r\simeq 1\,$pc), and its central
parts have condensed due to cooling.    

At the onset of the blue loop, the central star reaches close-to-critical 
rotation, and ejects a dense equatorial disk (Heger \& Langer 1998). While mass 
and time scales differ, this phenomenon occurs quite analogous to the
simulation of the outburst of $\eta\,$Carinae by Langer et al. (~{\cite{LGM99}}).
Like in this case, the ensuing BSG wind sweeps up the preceding
slow wind material into an ``hour glass'' structure
(Fig.~\ref{FigDenstemp}).
On a time scale of a few $10^4\,$yr, this hour glass expands
into the sphere defined by the RSG shell, with a maximum velocity
of $\sim$130$\,\kms$ (Fig.~\ref{FigVelshell}).  The faster polar
parts of the hour glass hit the inner edge of the RSG shell first.
The collision creates a hot ($T\simeq 10^5$\,K) and dense 
($n\simeq 10\,$cm$^{-3}$) pair of polar caps.
As time proceeds, the collision zone moves to lower latitudes
of the RSG shell and becomes more confined in latitude. 
At the same time, the interaction of the 
BSG wind with the equatorial disk defines a second, ring shaped collision
zone in the equatorial plane, which expands with time with a velocity
of 18$\,\kms$. 

Figure~\ref{FigEmis} shows snapshots of the emissivity map,
according to the employed cooling curve in our hydro simulations,
for three slices in time, along with projection maps constructed
from rotationally symmetric 3D-structures obtained from the 2D maps.
Here, only emission from gas in the temperature range between
$10^{4}$~K and $10^{6}$~K is considered, which is the dominant component. 
Hotter gas, which is formed from the reverse shock of the collision,
might be observable in the X-ray regime; the
peak luminosity of this component in our model is $10^{33}$\,erg\,s$^{-1}$,
which is achieved about 50\,000\,yr after the onset of the collision.
At an early interaction stage, the radiation is dominated by two 
polar caps and one equatorial ring, later on
by two mid-latitude rings and one fading smaller equatorial ring,
and finally two mid-latitude rings at rather low latitude are visible.
Those two rings gradually move to the equatorial plane while fading.
 The full time dependence
of the emission structure is shown in an accompanying movie which
is available from the A\&A website.

The energy budget for the collision of the polar caps of the hour glass
with the RSG shell follows directly from the stellar properties.
The polar caps have an emissivity of 
$l\simeq 10^{-21}\,$erg\,cm$^{-3}$\,s$^{-1}$ in a volume of 
$V \simeq 4\pi r^2 \Delta r = 4\times10^{54}\,$cm$^3$ (with $r=0.5\,$pc and $\Delta r = 0.04\,$pc;
see Figure~\ref{FigEmis}).
Thus, they shine with a total luminosity of $4\times10^{33}\,$erg\,s$^{-1}$, i.e.
roughly one solar luminosity, with a time scale of $\tau_{\rm rad}=l/u \simeq 9000\,$yr,
where $u={3\over 2} n k T$ is the internal energy of the gas, and $T\simeq 10^5\,$K
and $n\simeq 13\,$cm$^{-3}$ (corresponding to $\rho  \simeq 10^{-23}\,$g\,cm$^{-3}$;
Fig.~\ref{FigDenstemp}). The total radiated energy of the polar caps is about
$E_{\rm rad} \simeq \tau_{\rm rad} L \simeq 10^{45}\,$erg. This corresponds well
to the kinetic energy release due to the braking of the polar caps, which
reach their maximum velocity of $\varv \simeq 130\kms$ at the time of collision,
where it is reduced to $\varv \simeq 50\kms$ (Fig.~\ref{FigVelshell}). That is,
$\Delta E_{\rm kin} = {1\over 2} \Delta M \Delta \varv^2 \simeq 8\times10^{44}\,$erg, with
$\Delta M \simeq  1.2\times10^{-2}\mso$ and $\Delta \varv\simeq 80\kms$.
This kinetic energy can be compared with the BSG wind kinetic energy, which,
for $\dot M \simeq 10^{-6.8} \msoy$ and $\varv_{\rm wind}\simeq 300\kms$
(Fig.~\ref{FigMvvo}), yields
$\sim 1.2\times10^{45}\,$erg over a time period of 9000\,yr.
Thus, the polar caps shine because the hour glass shaped BSG shell 
collides with the spherical RSG shell.

A similar consideration could be made for the inner ring,
which is produced by the collision of the BSG wind with the equatorial disk ejected
by the central star during the phase of near critical rotation. 
The disk properties depend on the wind properties of the star during this
phase. However, in particular their latitude dependence, can not be expected to be reliably
predicted within the current assumptions. The total mass of the disk is determined by the
mass loss of the star at critical rotation.

\section{Discussion}

\begin{figure}[!h]
   \centering
   \resizebox{\hsize}{!}{\includegraphics[width=0.95
\textwidth]{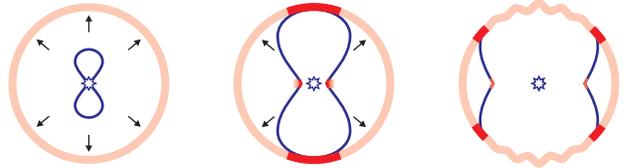}}
      \caption{ Schematic representation of the interaction of the hour glass shaped BSG shell
(blue) with the RSG shell (orange). The collision regions are marked with red color; they form the brightest
parts of the nebula.}
       \label{FigCartoon}
   \end{figure}

Figure~\ref{FigCartoon} illustrates a simplified picture of 
the formation of multiple ring nebulae, according to our model.
It contains two kinematic components: a stationary, spherical RSG shell
and an expanding hour glass structure. The strongly emitting parts of the structure are
the collision surfaces, which are marked in red in the figure.
We believe that both kinematic components are realized by nature in the
circumstellar medium of massive stars.
RSG shells are unambiguously predicted (Garc\'{i}a-Segura~{\cite{G07}}) 
and while they are not yet observationally confirmed, there seems
to be no way to avoid their formation. Expanding hour glass
structures, on the other hand, are a well documented feature in circumstellar nebulae
of low and high mass stars (see Nota et al.~{\cite{N95}}; Langer et al.~{\cite{LGM99}}) and are thought to be confined by a
circumstellar disk in the equatorial plane of the central star. 
 
A number of predictions emerge from this simple model. First, the collision starts
about 10$^4$\,yr after the onset of the blue loop. This timing is set by the
expansion speed of the BSG shell and the radius of the RSG shell. Second, the 
life time of the nebula is determined by the duration of the collision phase,
as the emission time scale is shorter than that. In our example, this is about
$10^4$\,yr, or about 1\% of the core helium burning life time. This provides an upper
limit to the expected number of triple ring nebulae. Third, the rotation rate of the central
star during the collision is high for a BSG, since it just about recovers from critical rotation.
At the time of maximum brightness of the nebula, the equatorial rotation rate of
our central star is about 80$\kms$ (Fig.~\ref{FigMvvo}). Fourth, as all the material in the 
nebula is ejected after the first dredge-up phase of the central star, the nebula material 
is nitrogen-rich, here enhanced by a factor of~6.5, and carbon and oxygen depleted 
by factors of~6.5 and~1.5, respectively. We note that the level of N-enrichment predicted by current stellar evolution models 
is quite uncertain (see Hunter et al. {\cite{Hu08}}), but a RSG phase is still expected to produce some nitrogen enhancement.
Due to the assumptions of efficient rotational mixing (Heger \& Langer~{\cite{HL00}), 
the star and nebula in our model are more 
enriched than expected from non-rotating stellar models.  
Fifth, one ingredient of our simple model, namely the RSG shell, is expected
for massive stars, but not so for low mass star which produce planetary nebulae.
Therefore, while quite analogous expanding hour glass structures are observed for both cases
(Langer~{\cite{L00}}), multiple ring nebulae formed through the collision
process shown in Fig.~\ref{FigCartoon} are expected around massive stars, but not as planetary nebulae. 
In this sense, the polar caps observed around the blue supergiant Sher~25
might be considered as the first indirect empirical confirmation of a RSG shell. 

Previous models of multiple ring nebulae were mostly constructed in the context
of the triple-ring structure observed around SN\,1987A (Burrows et al.~{\cite{B95}};
Crotts \& Heathcote~{\cite{CH00}}).
While single star models often fail to explain important features
(e.g., Martin \& Arnett~{\cite{MA95}}; 
Meyer~{\cite{Me97}}; Woosley et al.~{\cite{W97}}),
many invoke rather complex binary phenomena
(e.g., Podsiadlowski et al.~{\cite{P91}}; Blondin \& Lundqvist~{\cite{BL93}}; 
Llyod et al.~{\cite{L95}}; Morris \& Podsiadlowski~{\cite{PM05}}).
Whereas we do not attempt to reproduce the circumstellar medium of
SN~1987A, a single star approach with suitable
choices for the major parameters in our model (initial mass,
initial rotation rate, metallicity) appears promising and will be pursued in the near future.
The current failure of single star models to produce suitable blue loops
and blue supergiant pre-supernova models may have to do more with
missing physics in stellar evolution models rather than supporting
the evidence for a binary progenitor of SN\,1987A
(Woosley et al.~{\cite{W97}})

Various multiple ring nebulae around blue supergiants have been observed in the last
20~years (Smith et al.~{\cite{SBW07}}). While our generic numerical model was not designed to
correspond to any of them, many of the general properties of these nebulae are well
reproduced. Most striking is the agreement of the emission geometries. 
While the nebula around the B1.5~Ia supergiant Sher~25 shows two polar caps and one equatorial ring
(Brandner et al.~{\cite{BGCW97}}) and the other objects discussed by Smith et al.~{\cite{SBW07}}
rather show narrow rings, including the ``twin'' of the SN~1987A nebula around 
HD~168625 (Smith~{\cite{Sm07}})
all these structure occur as an evolutionary sequence 
in our model. Expansion velocities of the inner ring ($\sim 18\kms$) and the outer
collision products ($\sim 50\kms$), the spatial scale of about 1\,pc, and the
kinematic nebula age agree rather well with empirical values. The rotation velocity
of our stellar model fits well to the derived value of $\sim 70\kms$ for Sher~25
(Hendry et al.~{\cite{He08}}). Central star and nebula of our model are nitrogen enriched, 
as are most of the observed nebulae.

We note that the emission in our model is caused by compressional heating, 
which may be in conflict with evidence for photoionization being the dominant process in some
observed multiple ring nebulae (see Smith et al.~{\cite{SBW07}}). And indeed, looking at the density distributions
shown in Fig.~\ref{FigDenstemp}, which might resemble emission geometries in the pure
photo-ionization case, the situation appears more complex.
In our simulation, the thick RSG shell ($\Delta r \simeq r \simeq 1\,$pc)
collapses in two parts (at $r\simeq 1.2\,$ pc and $r\simeq 1.7\,$) due to a cooling
instability. However, this collapse is questionable since it requires a long timescale
--- our shell has an age of close to $10^6\,$yr, while in many cases the shell will
be much younger at the time of collision ---, and since the employed cooling function 
is uncertain for temperatures below $10\,000\,$K. 
Without this collapse, its density would be only
about $2\times10^{-25}\,$g\,cm$^{-3}$ (or 0.1 particles/cm$^3$), which may render it
unobservable even if it were photoionized. However, even in the case of the 
collapsed RSG shell as in our simulation, the collision leads to a clear density
enhancement. In panel~2 of Fig.~\ref{FigDenstemp}, we see that in our model,
the density enhancement in the polar caps is about a factor~5. This is to be considered
a lower limit, as higher resolution models might approach the theoretically expected
enhancement factor of about~100, which follows from the (well realized) isothermal
shock approximation and a Mach number of about~10. The lower panels of Fig.~\ref{FigDenstemp}
show that in order to represent the rings of SN~1987A, further refinements
are required, which, as we think, might be achieved by altering the properties
of the RSG shell. For this particular case, this may indeed be justified,
as the life time of the final RSG stage of the progenitor of SN~1987A
might have been quite short (Woosley~{\cite{W88}}, 
Langer~{\cite{L91a}}).

Despite that our model does not fit any of the observed cases in detail,
the approximate agreement with most general properties of this class of objects
encourages to produce tailored models for individual nebulae as next step.
Our results indicate that stars with multiple ring nebulae might just have
left the RSG branch --- as stellar evolution models argued for the case of SN~1987A
(Woosley~{\cite{W88}}, Langer~{\cite{L91a}}) and furthermore, that binarity may not be required 
to obtain multiple ring emission geometries. 

\begin{acknowledgements}

We are grateful to the anonymous referee for helpful remarks which 
lead to significant improvements of this paper.
We would like to thank Karolien Lefever for providing us with a grid
of atmospheric models for hot stars.
We would like to thank Anthony Marston, Nathan Smith and Bob van Veelen 
for helpful discussions. A.J.v.M. acknowledges support from NFS grant AST-0507581.
AH was supported by the DOE Program for Scientific Discovery
through Advanced Computing (SciDAC; grants DOEFC02-
01ER41176 and DOE-FC02-06ER41438) and performed this work under the
auspices of the National Nuclear Security Administration of the
U.S. Department of Energy at Los Alamos National Laboratory
under Contract No. DE-AC52-06NA25396.    
This work was supported by the Dutch Stichting Nationale Computerfaciliteiten 
(NCF).

\end{acknowledgements}

\Online
\begin{appendix}{A}

\section{Additional information}

\begin{figure}[!h]
   \centering
   \resizebox{\hsize}{!}{\includegraphics[width=0.95
\textwidth]{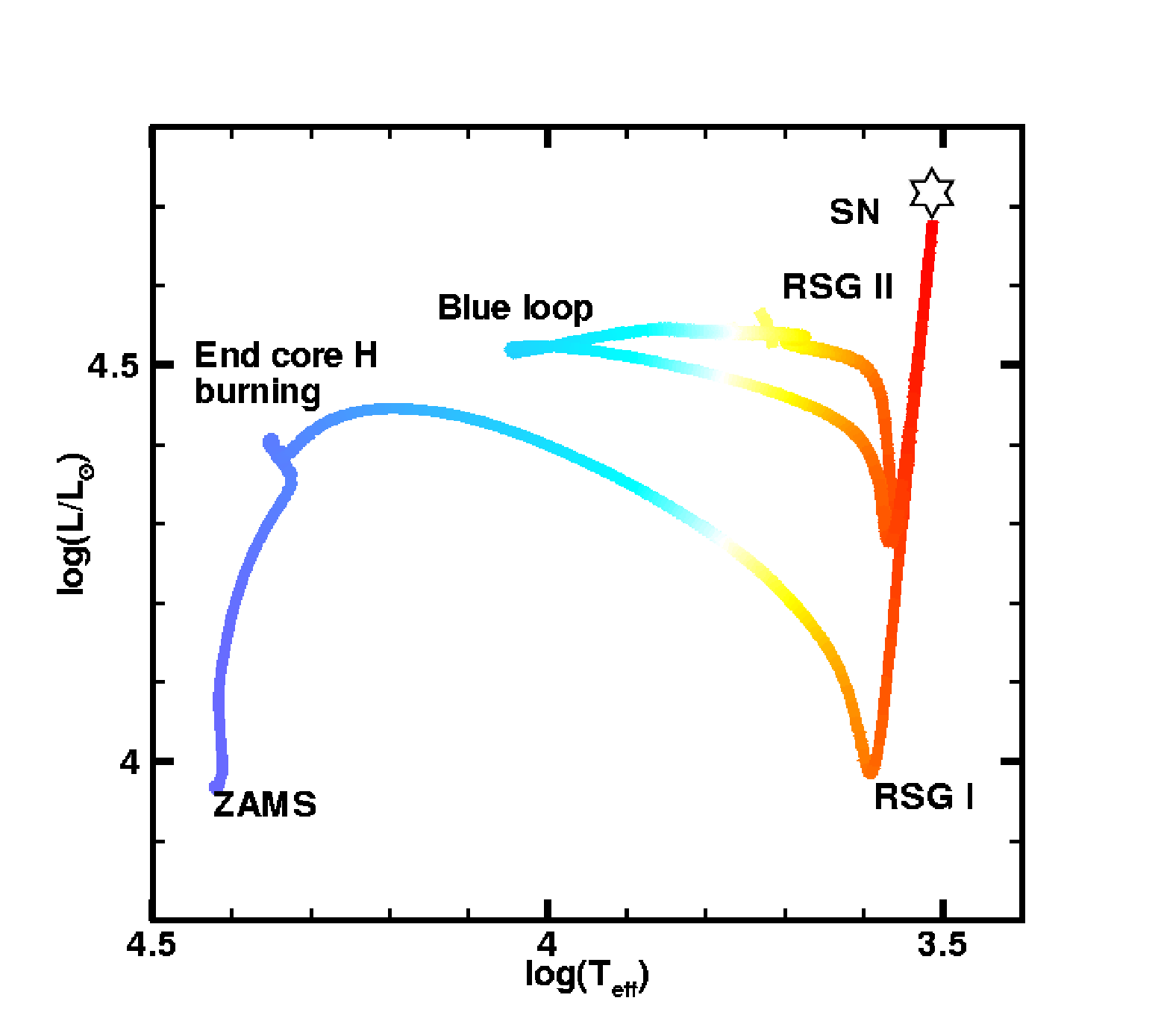}}
      \caption{ Evolution of our rotating 12~$\mso$ stellar evolution model
  in the Hertzsprung-Russell diagram, from the zero age main sequence to the pre-supernova
  stage. The initial metallicity of the model is $Z=0.02$, and the initial equatorial rotation 
  velocity is $328\kms$.}
       \label{FigHrd}
   \end{figure}

\begin{figure}[!h]
   \centering
   \resizebox{\hsize}{!}{\includegraphics[width=0.95 \textwidth]{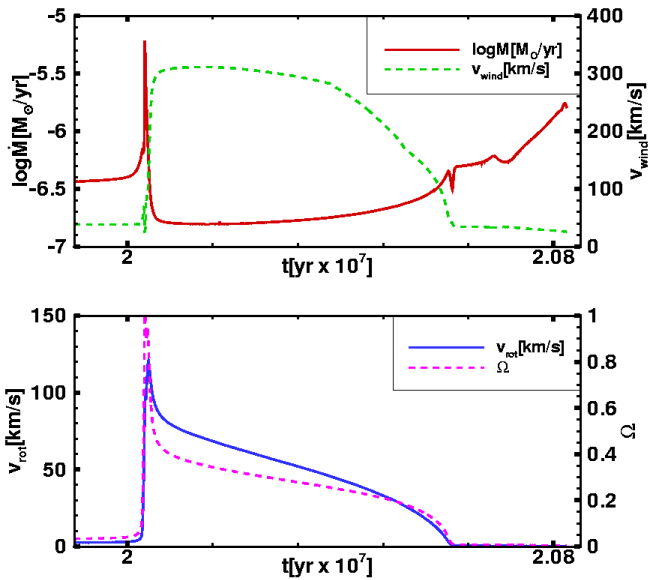}}
      \caption{Time dependence of various quantities of the employed 12$\mso$
  model, for about the final one million before core collapse.
  The upper panel shows 
  stellar mass loss rate and terminal wind velocity. The lower
      panel depicts stellar rotational velocity, and the ratio of stellar and
   critical rotation rate, $\Omega$.}
       \label{FigMvvo}
   \end{figure}

\begin{figure}[!h]
   \centering
   \resizebox{\hsize}{!}{\includegraphics[width=0.95
\textwidth]{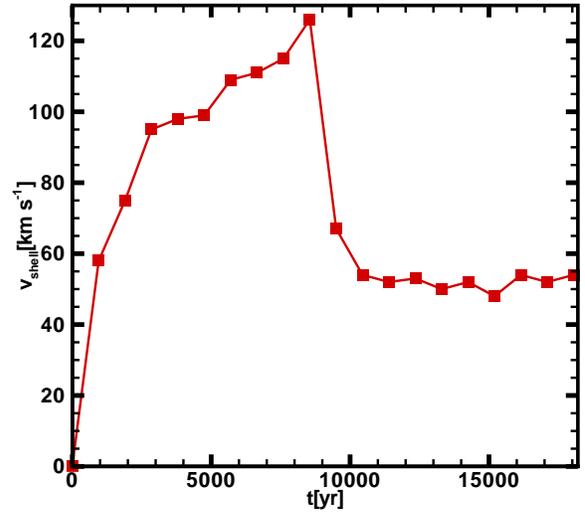}}
      \caption{Blue supergiant shell velocity, covering a timespan of 18\,000~yrs, 
      from its formation ($t=0$), until its collision with the red supergiant shell. 
      The time between two squares corresponds to 950~yrs. The time zero point corresponds to 
      $\sim 9000$~yrs before the time of the first snapshot in Fig.\ref{FigEmis}.}
       \label{FigVelshell}
   \end{figure}

\begin{figure*}[!h]
   \centering
   \includegraphics[width=0.65 \textwidth, angle=270]{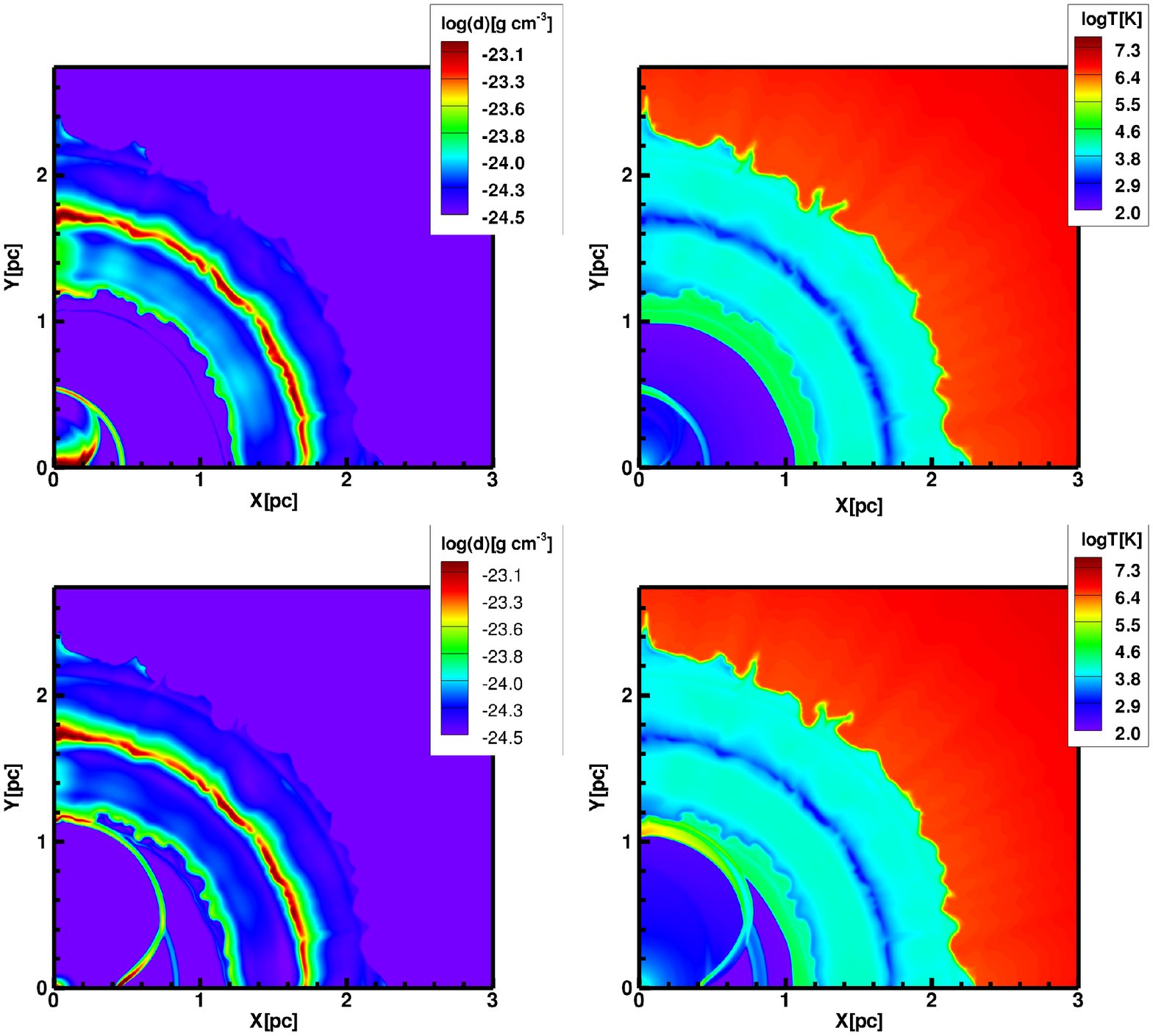}
   \includegraphics[width=0.65 \textwidth, angle=270]{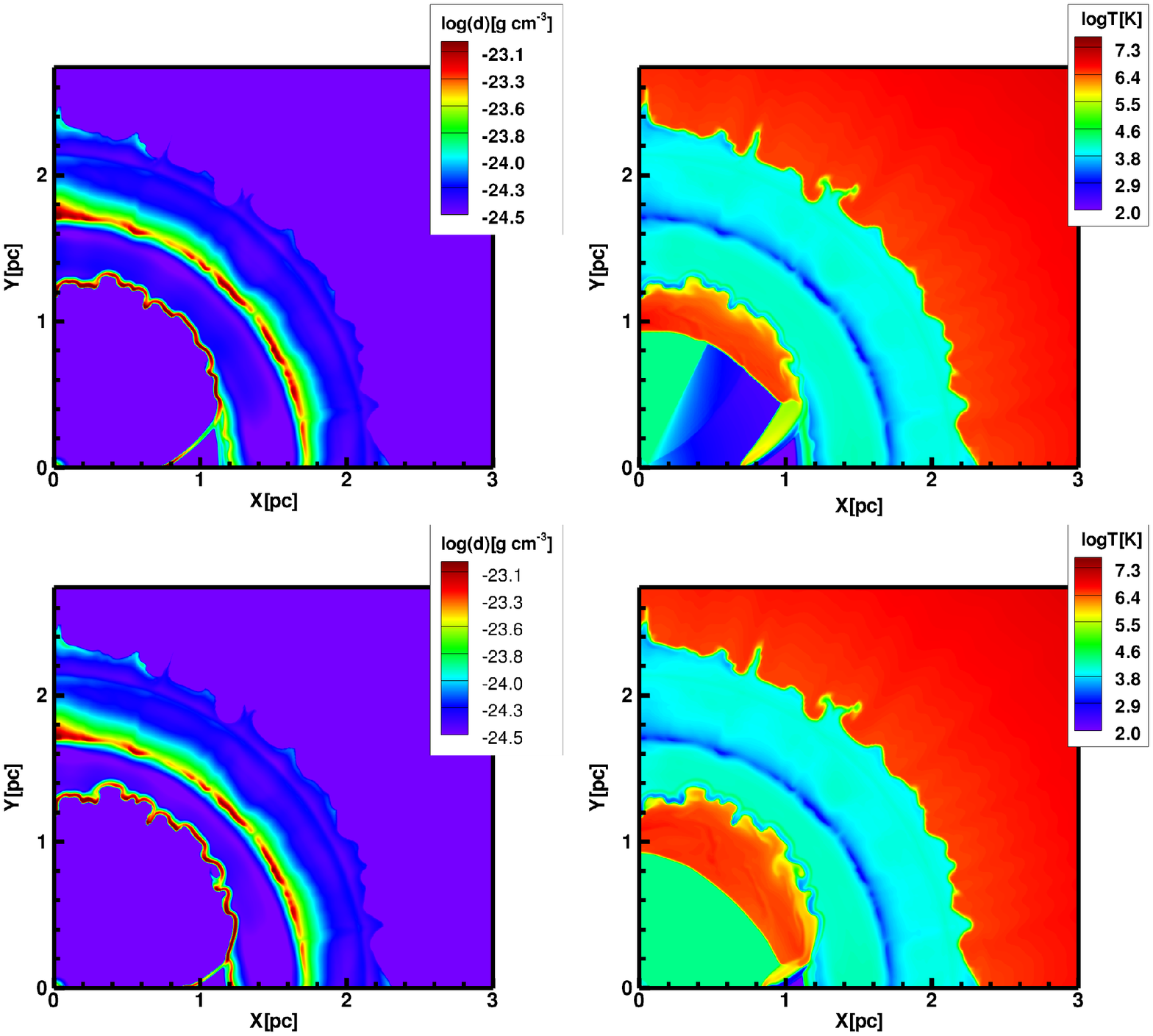}
      \caption{Snapshots of the density (left figures) and temperature (right figures) 
distributions of the
circumstellar material around our 12~$\mso$ model
after its first red supergiant stage. The
first panel corresponds to beginning of the blue supergiant stage.
The spherical red supergiant shell is
situated between 1~pc and 2.2~pc. The hourglass shaped blue
supergiant shell occupies the inner 0.4-0.5~pc.
The second panel depicts the situation 4500\,yr later, when the polar parts of the BSG shell collide with the RSG shell.
Panels~3 and~4 show the density and temperature distributions
10\,700\,yr and 13\,600\,yr after the first time of Panel~1.}
    \label{FigDenstemp}
   \end{figure*}

\begin{figure}
\caption{{\bf Movie description} 
Movie of the projections of the 3D emission obtained by assuming
rotational symmetry of the 2D structures obtained from the hydro
simulation, viewed with an inclination angle of 60$^{\circ}$
(cf. Fig.~\ref{FigEmis}, lower panel).
It covers a time span of 17000\,yr, starting at 2400~yrs after the time of the
first panel of Fig.~\ref{FigDenstemp}.
The time difference between two frames is about 475\,yr.}
\end{figure}

\end{appendix}

\end{document}